\documentclass[11pt]{article}
\usepackage[dvips]{graphicx}

\usepackage{epsfig}
\usepackage{graphicx}
\usepackage{indentfirst}
\usepackage{amsmath}
\usepackage{amsfonts}
\usepackage{amssymb}

\usepackage[pdftex,pdfmenubar=true,bookmarks=true,pdftoolbar=true]{hyperref}
\hypersetup{colorlinks,citecolor=green,filecolor=magenta,linkcolor=red,urlcolor=cyan,pdftex}
\begin{document}
\begin{center}
\begin{flushright}\begin{small}    
\end{small} \end{flushright} \vspace{1.5cm}
\huge{Inhomogeneous Universe in $f(T)$ Theory} 
\end{center}

\begin{center}
{\small  Manuel E. Rodrigues $^{(a,f,g)}$}\footnote{E-mail
address: esialg@gmail.com},
{\small  M. Hamani Daouda $^{(a,i)}$}\footnote{E-mail address:
daoudah8@yahoo.fr},
{\small    M. J. S. Houndjo $^{(b,c,h)}$}\footnote{E-mail address:
sthoundjo@yahoo.fr},
{\small    Ratbay Myrzakulov $^{(e)}$}\footnote{E-mail address:
rmyrzakulov@gmail.com} \,and
{\small    Muhammad Sharif $^{(d)}$}\footnote{E-mail address:
: msharif.math@pu.edu.pk}

\vskip 4mm

(a) \ Universidade Federal do Esp\'{\i}rito Santo \\
Centro de Ci\^{e}ncias
Exatas - Departamento de F\'{\i}sica\\
Av. Fernando Ferrari s/n - Campus de Goiabeiras\\ CEP 29075-910 -
Vit\'{o}ria/ES, Brazil \\
(b) \ Departamento de Ci\^{e}mcias Exatas - CEUNES\\
Universidade Federal do Esp\'irito Santo\\
CEP 29933-415 - S\~ao Mateus/ ES, Brazil\\
(c) \ Institut de Math´ematiques et de Sciences Physiques (IMSP)\\
01 BP 613 Porto-Novo, B\'enin\\
(d) Department of Mathematics, University of the Punjab,\\
Quaid-e-Azam Campus, Lahore-54590, Pakistan\\
(e) Eurasian International Center for Theoretical Physics\\
L.N. Gumilyov Eurasian National University, Astana 010008, Kazakhstan\\
(f)\ Faculdade de F\'{\i}sica, Universidade Federal do Par\'{a}, 66075-110, Bel\'em, Par\'{a}, Brazil\\
(g)\ Faculdade de Ci\^{e}ncias Exatas e Tecnologia, Universidade Federal do Par\'{a} - Campus Universit\'ario de Abaetetuba, CEP 68440-000, Abaetetuba, Par\'{a}, Brazil\\
(h)\  Facult\'e des Sciences et Techniques de Natitingou - Universit\'e de Parakou - B\'enin\\
(i)\ Department of Physics, Faculty of Sciences Abdou Moumouni University of Niamey, P.O. Box 10662,  NIAMEY-NIGER
\vskip 2mm
\end{center}

\begin{abstract}
\hspace{0,2cm}We obtain the equations of motions of the $f(T)$ theory considering the Lema\^itre-Tolman-Bondi's metric for a set of diagonal and non-diagonal tetrads. In the case of diagonal tetrads the equations of motion of the $f(T)$ theory impose a constant torsion or the same equations of the General Relativity, while in the case of non-diagonal set the equations are quite different from that obtained in GR. We show a simple example of an universe dominated by the matter for the two cases. The comparison of the mass in the non-diagonal case shows a sort of increased with respect to the diagonal one. We also perform two examples for the non-diagonal case. The first concerns a  black hole solution of type Sshwarzschild which presents a temperature higher than that of Schwarzschild, and a black hole in a dust-dominated universe.    
\end{abstract}
Pacs numbers: 04.50. Kd, 04.70.Bw, 04.20. Jb

\tableofcontents
\section{Introduction}
\hspace{0,2cm}A possible equivalence between the equations of General Relativity (GR)  can be obtained considering a space-time where the curvature contributions vanish and the unique non null contribution is that coming from the antisymmetric part of the connection. This is the scenario of the so-called Weitzenbok's space-time. Through this equivalence, we can analyse the physical phenomena of the Gravitation and the Cosmology, which until now are not clearly known using the GR. Hence, we can try to easily understand the contribution of the terms of higher order in curvature, added to the Einstein-Hilbert term as can be observed in the common theories, $f(R)$ \cite{odintsov,capozziello}, $f(G)$ \cite{sergei}, $f(R,\mathcal{T})$ \cite{sergei2} and $f(R,G)$  \cite{antoniadis}. In the case of the theory equivalent to the GR, the Teleparallel Theory (TT) \cite{pereira}, the attention is now attached to the torsion scalar $T$ which plays an important role in constructing the action of this theory. Hence, as in the case of GR, a generalization of the TT must contain terms of higher order in $T$ that we call $f(T)$ theory \cite{stephane2}, where $f(T)$ is an algebraic function of the torsion scalar $T$.
\par
Several works have been done in this sense these recent months in the $f(T)$ theory, with various interesting results \cite{x,MR}. However, there is not still more progress in introducing new symmetries, as in the case of TT. Therefore, in order to analyse the possible results we propose here to introduce a new symmetry, that of the Lema\^itre-Tolman's (LT) models. This may help us to understand more about the gravitational and cosmological phenomena cited above.
\par 
The solutions called inhomogeneous have various applications in the GR. A good revision of these applications is shown in \cite{krasinski}. A particular case of these solutions  is that of LT \cite{tolman}. The models have been used in redshift drift \cite{uzan}, CMB \cite{cmb}, interpretation of supernova observations \cite{SN}, averaging \cite{buchert}, formation of black holes \cite{fi}, of galaxy clusters \cite{krasinski1}, superclusters \cite{bolejko}, cosmic voids \cite{bolejko1} and collapse from the perspective of loop quantum gravity \cite{martin}.
\par
In this paper, we consider the symmetries of the Lema\^itre-Tolman's metric for comparing the physics of the solution coming from the $f(T)$ theory with the well known results in the GR. To do this, we take two workable choices of tetrads, the diagonal and the non-diagonal one, where we will get the real notion of the main different with respect to the GR.
\par
The paper is organized as follows. In Sec. \ref{sec2}, we explicitly present the equations of motion in $f(T)$ theory. The Sec. \ref{sec3} is devoted to the characterization of the geometry of an inhomogeneous universe. In Sec. \ref{sec4}, a set of non-diagonal tetrads for the metric of LTB is presented. In Sec \ref{sec5} we obtain new solutions, where we perform examples of a  dust-dominated universe in the subsection \ref{sec5.1}, a Schwarzschild-type black hole solution in the subsection \ref{sec5.2}, a  black hole solution in a dust-dominated universe in the subsection \ref{sec5.3} and other solutions in \ref{sec5.4}. The conclusion and perspective are presented in the Sec. \ref{sec6}. 
       

\section{\large The field equations from $f(T)$ theory}\label{sec2}

\hspace{0,2cm}In this section we will develop how obtaining the equations of motions for the $f(T)$ theory and the choice of matter model as an anisotropic fluid. 
\par
We start defining the line element as 
\begin{eqnarray}
dS^2=g_{\mu\nu}dx^{\mu}dx^{\nu}=\eta_{ab}\theta^{a}\theta^{b}\label{ele}\;,\\
\theta^{a}=e^{a}_{\;\;\mu}dx^{\mu}\;,\;dx^{\mu}=e_{a}^{\;\;\mu}\theta^{a}\label{the}\;,
\end{eqnarray}
where $g_{\mu\nu}$ is the metric of the space-time, $\eta_{ab}$ is the Minkowski's metric, $\theta^{a}$ are the tetrads and $e^{a}_{\;\;\mu}$ and their inverses $e_{a}^{\;\;\mu}$ are the tetrads matrices that satisfy $e^{a}_{\;\;\mu}e_{a}^{\;\;\nu}=\delta^{\nu}_{\mu}$ and  $e^{a}_{\;\;\mu}e_{b}^{\;\;\mu}=\delta^{a}_{b}$. The root of the determinant of the metric is given by $\sqrt{-g}=det[e^{a}_{\;\;\mu}]=e$. The Weitzenbok's connection is defined by 
\begin{eqnarray}
\Gamma^{\alpha}_{\mu\nu}=e_{i}^{\;\;\alpha}\partial_{\nu}e^{i}_{\;\;\mu}=-e^{i}_{\;\;\mu}\partial_{\nu}e_{i}^{\;\;\alpha}\label{co}\; .
\end{eqnarray}

Through the connection we can define the components of the torsion and the contorsion as
\begin{eqnarray}
T^{\alpha}_{\;\;\mu\nu}&=&\Gamma^{\alpha}_{\nu\mu}-\Gamma^{\alpha}_{\mu\nu}=e_{i}^{\;\;\alpha}\left(\partial_{\mu} e^{i}_{\;\;\nu}-\partial_{\nu} e^{i}_{\;\;\mu}\right)\label{tor}\;,\\
K^{\mu\nu}_{\;\;\;\;\alpha}&=&-\frac{1}{2}\left(T^{\mu\nu}_{\;\;\;\;\alpha}-T^{\nu\mu}_{\;\;\;\;\alpha}-T_{\alpha}^{\;\;\mu\nu}\right)\label{cont}\; .
\end{eqnarray}

For facilitating the description of the Lagrangian and the equations of motion, we can define another tensor from the components of torsion and the contorsion as
\begin{eqnarray}
S_{\alpha}^{\;\;\mu\nu}=\frac{1}{2}\left( K_{\;\;\;\;\alpha}^{\mu\nu}+\delta^{\mu}_{\alpha}T^{\beta\nu}_{\;\;\;\;\beta}-\delta^{\nu}_{\alpha}T^{\beta\mu}_{\;\;\;\;\beta}\right)\label{s}\;.
\end{eqnarray}
Now, defining the torsion scalar 
\begin{equation}
T=T^{\alpha}_{\;\;\mu\nu}S_{\alpha}^{\;\;\mu\nu}\label{t1}\,\,\,,
\end{equation}
one can define the Lagrangian of the $f(T)$ theory, coupled with the matter as follows 
\begin{equation}
\mathcal{L}=ef(T)+\mathcal{L}_{Matter}\;.\label{lagran}
\end{equation}
The principle of least action leads to the Euler-Lagrange's equations. In order to use these equations we first write the quantities
\begin{eqnarray}
\frac{\partial\mathcal{L}}{\partial e^{a}_{\;\;\mu}}&=&f(T)ee_{a}^{\;\;\mu}+ef_{T}(T)4e_{a}^{\;\;\alpha}T^{\sigma}_{\;\;\nu\alpha}S_{\sigma}^{\;\;\mu\nu}+\frac{\partial\mathcal{L}_{Matter}}{\partial e^{a}_{\;\;\mu}}\;,\label{1}
\end{eqnarray}
\begin{eqnarray}
\partial_{\alpha}\left[\frac{\partial \mathcal{L}}{\partial (\partial_{\alpha}e^{a}_{\;\;\mu})}\right]=-4f_{T}(T)\partial_{\alpha}\left(ee_{a}^{\;\;\sigma}S_{\sigma}^{\;\;\mu\nu}\right)-4ee_{a}^{\;\;\sigma}S_{\sigma}^{\;\;\mu\alpha}\partial_{\alpha}T\,f_{TT}(T)+\partial_{\alpha}\left[\frac{\partial \mathcal{L}_{Matter}}{\partial (\partial_{\alpha}e^{a}_{\;\;\mu})}\right]\label{2}\;,
\end{eqnarray}
where $f_{T}(T)=df(T)/dT$ and $f_{TT}(T)=d^2f(T)/dT^2$. The equations of Euler-Lagrange are given by
\begin{eqnarray}\label{EL}
\frac{\partial\mathcal{L}}{\partial e^{a}_{\;\;\mu}}-\partial_{\alpha}\left[\frac{\partial \mathcal{L}}{\partial (\partial_{\alpha}e^{a}_{\;\;\mu})}\right]=0\;,
\end{eqnarray}
which, multiplying by $e^{-1}e^{a}_{\;\;\beta}/4$, yields 
\begin{eqnarray}
S_{\beta}^{\;\;\mu\alpha}\partial_{\alpha}T\,f_{TT}(T)+\left[e^{-1}e^{a}_{\;\;\beta}\partial_{\alpha}\left(ee_{a}^{\;\;\sigma}S_{\sigma}^{\;\;\mu\alpha}\right)+T^{\sigma}_{\;\;\nu\beta}S_{\sigma}^{\;\;\mu\nu}\right]f_{T}(T)+\frac{1}{4}\delta^{\mu}_{\beta}f(T)=4\pi \mathcal{T}^{\mu}_{\beta}\label{em}\;,
\end{eqnarray}
where the energy momentum tensor is given by 
\begin{eqnarray}
\mathcal{T}^{\mu}_{\beta}=-\frac{e^{-1}e^{a}_{\;\;\beta}}{16\pi}\left\{ \frac{\partial \mathcal{L}_{Matter}}{\partial e^{a}_{\;\;\mu}}-\partial_{\alpha}\left[\frac{\partial \mathcal{L}_{Matter}}{\partial (\partial_{\alpha}e^{a}_{\;\;\mu})}\right]\right\}\;.
\end{eqnarray}
For an anisotropic fluid, the energy momentum tensor is given by the expression
\begin{eqnarray}
\mathcal{T}^{\,\mu}_{\beta}=\left(\rho+p_t\right)u_{\beta}u^{\mu}-p_t \delta^{\mu}_{\beta}+\left(p_r-p_t\right)v_{\beta}v^{\mu}\label{tme}\; ,
\end{eqnarray}
where $u^{\mu}$ is the four-velocity, $v^{\mu}$ the unit space-like vector in the radial direction, $\rho$ the energy density, $p_r$ the pressure in the direction of $v^{\mu}$ (radial pressure) and $p_t$  the pressure orthogonal to $v_\mu$ (tangential pressure). Since we are assuming an anisotropic spherically symmetric matter, one has $p_r \neq p_t$, such that their equality corresponds to an isotropic fluid sphere.\par
An important point to be put out here is the non-invariance feature of this modified theory of gravity. We recall that any gravitational theory built from the metric and the therein geometrical quantities will always be Lorentz scalars and therefore must be invariant under local Lorentz transformations. However, in general, when dealing with theory based on the torsion this invariance does not hold. Let us show this in an explicit way by first considering the Lorentz transformation of the tetrad, as
\begin{eqnarray}
e^{a}_{\;\;\mu}\mapsto \Lambda ^{a}_{\;\;b}e^{b}_{\;\;\mu},
\end{eqnarray}
where $\Lambda^{a}_{\;\;b}$ denotes the local Lorentz transformation which satisfies
\begin{eqnarray}
\eta_{ac}\Lambda^{a}_{\;\;b}\Lambda^{c}_{\;\;d}=\eta_{bd}. 
\end{eqnarray}
By making use of this tetrad transformation, on can perform the torsion tensor (\ref{tor}) as
\begin{eqnarray}
T^{\alpha}_{\;\;\mu\nu}\mapsto T^{\alpha}_{\;\;\mu\nu}+\Lambda_{a}^{\;\;b}e_{b}^{\;\;\alpha}\left(e^{c}_{\;\;\nu}\partial_{\mu}\Lambda^{a}_{\;\;c}-e^{c}_{\;\;\mu}\partial_{\mu}\Lambda^{a}_{\;\;c}\right).
\end{eqnarray}
Thus, it is straightforward to see that for  a linear form of $f(T)$, i.e, teleparallel theory of gravity, there is invariance. However, for an arbitrary expression of the algebraic function $f(T)$, the modified $f(T)$ theories are not invariant under local Lorentz transformations \cite{3boe,4boe}.  Consequently, non-invariant theory is sensitive to the choice of the tetrad and depending to the set of tetrads, being diagonal or non-diagonal, different results may be obtained. Hence, the choice of tetrad is crucial in order to deflect the theory from the teleparallel one. This does not invalidate this generalization of the GR, but it can be shown that the newtonian limit is obtained for a particular case \cite{stephane2}.


\section{\large  The geometry of an inhomogeneous universe}\label{sec3}

\hspace{0,2cm}Given the metric of Lema\^itre-Tolman-Bondi \cite{tolman,bondi}
\begin{equation}
dS^2=dt^2-B^2(r,t)dr^2-A^{2} (r,t)\left(d\theta^{2}+\sin^{2}\left(\theta\right)d\phi^{2}\right)
\label{ltb}\;,
\end{equation}
we can describe this space-time through the following set of diagonal tetrads
\begin{eqnarray}
\left\{e^{a}_{\;\;\mu}\right\}=diag\left[ 1\,,\,B(r,t)\,,\,A(r,t)\,,\,A(r,t)\sin\theta\right]\,,\\
\left\{e_{a}^{\;\;\mu}\right\}=diag\left[ 1\,,\,B^{-1}(r,t)\,,\,A^{-1}(r,t)\,,\,A^{-1}(r,t)\sin^{-1}\theta\right]\;,
\end{eqnarray}
where we define the determinant of the tetrads by  $e=det[e^{a}_{\;\;\mu}]=A^2B\sin\theta$. The non null components of the torsion  (\ref{tor}) are 
\begin{eqnarray}\label{ctor}
\left\{\begin{array}{llllll}
T^{1}_{\;\;01}=-T^{1}_{\;\;10}=e_{1}^{\;\;1}\partial_{0}e^{1}_{\;\;1}=B^{-1}\dot{B},\\
T^{2}_{\;\;12}=-T^{2}_{\;\;21}=e_{2}^{\;\;2}\partial_{1}e^{2}_{\;\;2}=A^{-1}A^{\prime},\\
T^{3}_{\;\;13}=-T^{3}_{\;\;31}=e_{3}^{\;\;3}\partial_{1}e^{3}_{\;\;3}=A^{-1}A^{\prime},\\
T^{2}_{\;\;02}=-T^{2}_{\;\;20}=e_{2}^{\;\;2}\partial_{0}e^{2}_{\;\;2}=A^{-1}\dot{A},\\
T^{3}_{\;\;03}=-T^{3}_{\;\;30}=e_{3}^{\;\;3}\partial_{0}e^{3}_{\;\;3}=A^{-1}\dot{A},\\
T^{3}_{\;\;23}=-T^{3}_{\;\;32}=e_{3}^{\;\;3}\partial_{2}e^{3}_{\;\;3}=\cot\theta,
\end{array}\right.
\end{eqnarray}
where the ``dot" indicates the derivative with respect to the time  $t$ and the ``prime" the derivative with respect to the radial coordinate  $r$. The non null components of the contorsion  (\ref{cont}) are 
\begin{eqnarray}\label{ccont}
\left\{\begin{array}{llllll}
K^{01}_{\;\;\;\;1}=-K^{10}_{\;\;\;\;1}=g^{00}T^{1}_{\;\;01}=B^{-1}\dot{B},\\
K^{02}_{\;\;\;\;2}=-K^{20}_{\;\;\;\;2}=g^{00}T^{2}_{\;\;02}=A^{-1}\dot{A},\\
K^{03}_{\;\;\;\;3}=-K^{30}_{\;\;\;\;3}=g^{00}T^{3}_{\;\;03}=A^{-1}\dot{A},\\
K^{12}_{\;\;\;\;2}=-K^{21}_{\;\;\;\;2}=g^{11}T^{2}_{\;\;12}=-\frac{A^{\prime}}{AB^2},\\
K^{13}_{\;\;\;\;3}=-K^{31}_{\;\;\;\;3}=g^{11}T^{3}_{\;\;13}=-\frac{A^{\prime}}{AB^2},\\
K^{23}_{\;\;\;\;3}=-K^{32}_{\;\;\;\;3}=g^{22}T^{3}_{\;\;23}=-\frac{\cot\theta}{A^2}\;.
\end{array}\right.
\end{eqnarray}
We can now calculate the non null components of the tensor  $S_{\alpha}^{\;\;\mu\nu}$ in (\ref{s}), which are given by 
\begin{eqnarray}\label{cs}
\left\{\begin{array}{llllll}
S_{0}^{\;\;10}=-S_{0}^{\;\;01}=-\frac{1}{2}g^{11}T^{\beta}_{\;\;1\beta}=\frac{A^{\prime}}{AB^2},\\
S_{0}^{\;\;20}=-S_{0}^{\;\;02}=-\frac{1}{2}g^{22}T^{\beta}_{\;\;2\beta}=\frac{\cot\theta}{2A^2},\\
S_{1}^{\;\;01}=-S_{1}^{\;\;10}=\frac{1}{2}\left(K^{01}_{\;\;\;\;1}-T^{\beta 0}_{\;\;\;\;\beta}\right)=-A^{-1}\dot{A},\\
S_{1}^{\;\;21}=-S_{1}^{\;\;12}=-\frac{1}{2}g^{22}T^{\beta}_{\;\;2\beta}=\frac{\cot\theta}{2A^2},\\
S_{2}^{\;\;02}=-S_{2}^{\;\;20}=\frac{1}{2}\left(K^{02}_{\;\;\;\;2}-T^{\beta 0}_{\;\;\;\;\beta}\right)=-\frac{1}{2}\left(A^{-1}\dot{A}+B^{-1}\dot{B}\right),\\
S_{2}^{\;\;12}=-S_{2}^{\;\;21}=\frac{1}{2}\left(K^{12}_{\;\;\;\;2}-T^{\beta 1}_{\;\;\;\;\beta}\right)=\frac{A^{\prime}}{2AB^2},\\
S_{3}^{\;\;03}=-S_{3}^{\;\;30}=\frac{1}{2}\left(K^{03}_{\;\;\;\;3}-T^{\beta 0}_{\;\;\;\;\beta}\right)=-\frac{1}{2}\left(A^{-1}\dot{A}+B^{-1}\dot{B}\right),\\
S_{3}^{\;\;13}=-S_{3}^{\;\;31}=\frac{1}{2}\left(K^{13}_{\;\;\;\;3}-T^{\beta 1}_{\;\;\;\;\beta}\right)=\frac{A^{\prime}}{2AB^2}\;.
\end{array}\right.
\end{eqnarray}
Through the definition of the torsion scalar (\ref{t1}) and of the components (\ref{ctor}) and  (\ref{cs}), one obtains 
\begin{equation}
T=2\left[\left(\frac{A^{\prime}}{AB}\right)^2-2\frac{\dot{A}\dot{B}}{AB}-\left(\frac{\dot{A}}{A}\right)^2\right]\label{te}\;.
\end{equation}

We now obtain two equations that impose constraints to the $f(T)$ theory such that it becomes equivalent to the TT, which is the case where the algebraic function $f(T)$ is a linear function of the torsion scalar $T$. For the first of them, it is sufficient to put $\beta=0$ and $\mu=2$ in (\ref{em}), which leads to 
\begin{equation}
\frac{\cot\theta}{2A^2}\dot{T}f_{TT}(T)=0\label{imp1}\;,
\end{equation}
and for the second, it can be just put $\beta=1$ and $\mu=2$ in (\ref{em}), which yields  
\begin{equation}
\frac{\cot\theta}{2A^2}T^{\prime}f_{TT}(T)=0\label{imp2}\;.
\end{equation}
The equations (\ref{imp1}) and (\ref{imp2}) inform that, or the torsion scalar is a constant, which does not yield any interesting result for the metric of LTB, or the algebraic function $f(T)$ is linear in $T$. An imposition of the form (\ref{imp2}) has been obtained for a set of diagonal tetrads in the case of a spherically symmetric and static metric \cite{stephane1,stephane2}. In the next section, we will see the case where it will be taken a set of non-diagonal tetrads for describing the metric of LTB (\ref{ltb}).  
\par
Substituting the components (\ref{ctor})-(\ref{cs}) and the torsion scalar (\ref{te}) in (\ref{em}), for the case of the components  $0-0$, $1-1$ and  $2-2$, we get the following equations of motion 
\begin{eqnarray}\label{em2}
\left\{\begin{array}{lll}
8\pi \rho=-2\frac{A^{\prime\prime}}{AB^2}+2\frac{A^{\prime}B^{\prime}}{AB^3}+2\frac{\dot{A}\dot{B}}{AB}+\frac{1}{A^2}+\left(\frac{\dot{A}}{A}\right)^2-\left(\frac{A^{\prime}}{AB}\right)^2\;,\\
-8\pi p_{r}=2\frac{\ddot{A}}{A}+\frac{1}{A^2}+\left(\frac{\dot{A}}{A}\right)^2-\left(\frac{A^{\prime}}{AB}\right)^2\;,\\
-8\pi p_{t}=-\frac{A^{\prime\prime}}{AB^2}+\frac{\ddot{A}}{A}+\frac{\ddot{B}}{B}+\frac{\dot{A}\dot{B}}{AB}+\frac{A^{\prime}B^{\prime}}{AB^3}\;,
\end{array}\right.
\end{eqnarray}
that are identical to the equations of GR \cite{grande}. This is not surprising since the TT theory is dynamically equivalent to the GR \cite{barrow1}. The famous symmetry of the metric of LTB is recuperated when one takes $\beta=0$ and  $\mu=1$ in (\ref{em})
\begin{equation}
\left(\dot{A}^{\prime}B-A^{\prime}\dot{B}\right)f_{T}(T)=0\;,
\end{equation}
which, after integration leads to 
\begin{equation}
B(r,t)=c^{-1}(r)A^{\prime}(r,t)\label{b1}\;.
\end{equation}

The function that appears in  (\ref{b1}) as integration constant for the coordinate $t$ can be fixed, as in the case of GR, due to its relationship with the spatial curvature $c(r)=\sqrt{1-k(r)}$ \cite{grande}. 
\par
A direct application of these results is the so-called limit of the Friedmann-Lema\^itre-Robertson-Walker's universe. To do this, let us consider the equation of conservation for an energy momentum tensor with null radial and tangential pressures ($p_r=p_t=0$ in (\ref{tme})):
\begin{eqnarray}\label{ec}
\nabla_{\mu}\mathcal{T}^{\mu\nu}=0\;,
\end{eqnarray}
such that, using (\ref{b1}) and integrating (\ref{ec}), one gets
\begin{eqnarray}
\rho(r,t)=\frac{c(r)\rho_{0}(r)}{A^2(r,t)A^{\prime}(r,t)}\label{dens1}\;,
\end{eqnarray}
where $\rho_{0}(r)$ is an algebraic function of $r$, coming from the integration in $t$. The first equation in  (\ref{em2}) can be rewritten as 
\begin{eqnarray}\label{eq1}
8\pi\rho A^2A^{\prime}=\left[A\left(1-c^2(r)+\dot{A}^2\right)\right]^{\prime}\;.
\end{eqnarray}
In this diagonal case, by defining the mass of a black hole as  
\begin{eqnarray}\label{mass}
M_{D}(r)=8\pi\int^{r}_{0}c(y)\rho(y,t)A^2(y,t)B(y,t)dy=8\pi\int^{r}_{0}c(y)\rho_{0}(r)dy\;,
\end{eqnarray}
the equation (\ref{eq1}) can be integrated, yielding 
\begin{eqnarray}
\dot{A}^2=\frac{M_{D}(r)}{A}+c^2(r)-1\label{eq2}\;.
\end{eqnarray}

A particular case of this equation is when $c(r)=1$, where the integration leads to 
\begin{eqnarray}
A(r,t)=\left[d(r)+\frac{3}{2}\sqrt{M_{D}(r)}\,t\right]^{2/3}\label{a1}\;,
\end{eqnarray}
where $d(r)$ is an algebraic function of $r$. Putting $d(r)=r^{3/2}$ and $M_{D} (r)=8\pi\bar{\rho}_{0}r^3/3$, where  $\bar{\rho}_{0}$ is a constant, we re-obtain the equations of Friedmann for an universe dominated by the matter \cite{gao}, where $A(r,t)=ra(t)$, $a(t)=[1+\sqrt{6\pi\bar{\rho}_{0}}t]^{2/3}$, $\rho(t)=\bar{\rho}_{0}/a^3(t)$ and $t_{B}=-(6\pi\bar{\rho}_{0})^{-1/2}$ representing the  Big Bang.  
\par
In the next section we will perform the calculus about the equations of motion for a set of non-diagonal tetrads.
\section{\large A set of non-diagonal tetrad}\label{sec4}

\hspace{0,2cm}We can also project in the tangent space to the LTB's metric (\ref{ltb}) through a set of non-diagonal tetrads  as follows 
\begin{eqnarray}\label{nontet}
\{e^{a}_{\;\;\mu}\}=\left[\begin{array}{cccc}
1&0&0&0\\
0&B(r,t)\sin\theta\cos\phi &A(r,t)\cos\theta\cos\phi &-A(r,t)\sin\theta\sin\phi\\
0&B(r,t)\sin\theta\sin\phi &A(r,t)\cos\theta\sin\phi &A(r,t)\sin\theta\cos\phi \\
0&B(r,t)\cos\theta &-A(r,t)\sin\theta &0
\end{array}\right]\;,
\end{eqnarray}  
whose inverse is 
\begin{eqnarray}\label{invnontet}
\{e_{a}^{\;\;\mu}\}=\left[\begin{array}{cccc}
1&0&0&0\\
0&B^{-1}(r,t)\sin\theta\cos\phi &B^{-1}(r,t)\sin\theta\sin\phi &B^{-1}(r,t)\cos\theta\\
0&A^{-1}(r,t)\cos\theta\sin\phi &A^{-1}(r,t)\cos\theta\sin\phi &-A^{-1}(r,t)\sin\theta \\
0&-A^{-1}(r,t)\sin^{-1}\theta\sin\phi &A^{-1}(r,t)\sin^{-1}\theta\cos\phi &0
\end{array}\right]\;.
\end{eqnarray}
  
By using (\ref{nontet}) and (\ref{invnontet}), the non null components of the torsion (\ref{tor}) are calculated as:
\begin{eqnarray}\label{ctnon}
\left\{\begin{array}{l}
T^{1}_{\;\;01}=B^{-1}\dot{B}\,,\,T^{2}_{\;\;02}=T^{3}_{\;\;03}=A^{-1}\dot{A}\,,\,T^{2}_{\;\;12}=T^{3}_{\;\;13}=-\frac{B-A^{\prime}}{A}\;.
\end{array}\right.
\end{eqnarray}
The non null components of the contorsion are
\begin{eqnarray}\label{cctnon}
\left\{\begin{array}{l}
K^{01}_{\;\;\;\;1}=B^{-1}\dot{B}\,,\,K^{02}_{\;\;\;2}=K^{03}_{\;\;\;\;3}=A^{-1}\dot{A}\,,\,
K^{12}_{\;\;\;\;2}=K^{13}_{\;\;\;\;3}=\frac{B-A^{\prime}}{AB^{2}}\;.
\end{array}\right.
\end{eqnarray}
The non null components of $S_{\alpha}^{\;\;\mu\nu}$ are
\begin{eqnarray}\label{csnon}
\left\{\begin{array}{l}
S_{0}^{\;\;01}=2S_{2}^{\;\;21}=2S_{3}^{\;\;31}=\frac{B-A^{\prime}}{AB^2}\,,\,S_{1}^{\;\;10}=A^{-1}\dot{A}\,,\,
S_{2}^{\;\;20}=S_{3}^{\;\;30}=\frac{1}{2}(A^{-1}\dot{A}+B^{-1}\dot{B})\;.
\end{array}\right.
\end{eqnarray}
Taking into account the components  (\ref{ctnon}) and  (\ref{csnon}), the torsion scalar (\ref{t1}) becomes 
\begin{eqnarray}\label{tenon}
T=-2\left[2\frac{\dot{A}\dot{B}}{AB}+\left(\frac{\dot{A}}{A}\right)^2-\frac{1}{A^2}+\frac{2A^{\prime}}{A^2B}-\left(\frac{A^{\prime}}{AB}\right)^2\right]\;.
\end{eqnarray}

Through (\ref{nontet})-(\ref{tenon}), the equations of motion  (\ref{em}) are given by 
\begin{eqnarray}\label{emnon}
\left\{\begin{array}{llll}
\frac{(B-A^{\prime})}{AB^2}T^{\prime}f_{TT}+f_{T}\left[\frac{T}{2}+\frac{1}{A^2}+4\frac{\dot{A}\dot{B}}{AB}+\frac{A^{\prime}B^{\prime}}{AB^3}-\frac{A^{\prime}}{A^2B}-\frac{A^{\prime\prime}}{AB^2}\right]+\frac{f}{4}=4\pi\rho\,,\\
-\frac{(B-A^{\prime})}{AB^2}\dot{T}f_{TT}+\frac{f_{T}}{AB}\left(A^{\prime}\dot{B}-\dot{A}^{\prime}B\right)=0\,,\\
-\frac{\dot{A}}{A}T^{\prime}f_{TT}-\frac{f_{T}}{AB^3}\left(A^{\prime}\dot{B}-\dot{A}^{\prime}B\right)=0\,,\\
\frac{\dot{A}}{A}T^{\prime}f_{TT}+f_{T}\left[\frac{T}{2}+\frac{\ddot{A}}{A}+3\frac{\dot{A}\dot{B}}{AB}+\frac{1}{A^2}-\frac{A^{\prime}}{A^2B}\right]+\frac{f}{4}=-4\pi p_{r}\,,\\
\frac{1}{2}\left[\left(\frac{\dot{A}}{A}+\frac{\dot{B}}{B}\right)\dot{T}+\frac{B-A^{\prime}}{AB^2}T^{\prime}\right]f_{TT}+\frac{f_{T}}{2}\left[\frac{T}{2}+\frac{\ddot{A}}{A}+\frac{\ddot{B}}{B}+5\frac{\dot{A}\dot{B}}{AB}+\frac{A^{\prime}B^{\prime}}{AB^3}-\frac{A^{\prime\prime}}{AB^2}\right]+\frac{f}{4}=-4\pi p_{t}\;.
\end{array}\right.
\end{eqnarray}
The equations of motion of the non-diagonal case are quite  different from the previous ones of the diagonal case. This is the proof of the dependence on the frame in $f(T)$ theory \cite{yapiskan}. In such a situation, the physics that results from this set of equations must present new data that may be important in understanding the subjects that still require attention in Cosmology and Astrophysics.

\section{New solutions}\label{sec5}

\subsection{Dust-dominated universe}\label{sec5.1}
Le us make a simplified example of the analysis of these equations. Taking the linear case $f(T)=T$, the second and third equations of (\ref{emnon}) yield again the constraint (\ref{b1}) well known in the metric of LTB. In order to compare with the example taken in the diagonal case, we put $p_r=p_t=0$ in (\ref{tme}) and rewrite the first equation of (\ref{emnon}) as
\begin{eqnarray}
8\pi\rho A^2A^{\prime}=8\pi\rho_{D}A^2A^{\prime}+2A^{\prime}\left[2c(r)-1-c^2(r)+\dot{A}^2\right]\;,\label{dens2}
\end{eqnarray}
where $\rho_{D}$ is the energy density in the diagonal case  in (\ref{eq1}). Using the equation of conservation  (\ref{ec}), we re-obtain  (\ref{dens1}), that suggests the same definition (\ref{mass}) for the mass in the non-diagonal case. But here we have the following identity:
\begin{eqnarray}
M(r,t)&=&8\pi\int^{r}_{0}c(y)\rho(y,t)A^2(y,t)B(y,t)dy=M_{D}(r)\nonumber\\
&&+2\int^{r}_{0}A^{\prime}(y,t)\left[2c(y)-1-c^2(y)+\dot{A}^2(y,t)\right]dy\label{mass2}\;.
\end{eqnarray}
Here, in this non-diagonal case, the mass of a black hole can depend on the time in general. Comparing with the diagonal case, it appears that the mass {\bf in} the non-diagonal case possesses an increased (decreased) due to the non-diagonal description of the matrix of the tetrads in  (\ref{nontet}). 
\par
Now, let us look at the particular case $c(r)=1$ and, as in the diagonal case, one supposes $A(r,t)=ra(t)$, with  $a(t)=[1+\sqrt{6\pi\bar{\rho}_{0}}t]^{2/3}$. Since $M_{D}(r)=A\dot{A}^2=8\pi\bar{\rho}_{0}r^3/3$, from (\ref{mass2}), one gets the following identity    
\begin{eqnarray}
M(r)=M_{D}(r)+\frac{2}{3}M_{D}(r)\;.\label{mass3}
\end{eqnarray}

This shows us that a significant value ($2/3$) of the mass (or energy density)     in the diagonal case is increased as contribution in the non-diagonal case. Note that the addition comes from the contribution of the set of the off-diagonal terms of the non-diagonal tetrads matrix.


\subsection{Black hole solution}\label{sec5.2}
For an exterior solution, in the vacuum, we get $\rho=0$ in the first equation of  (\ref{emnon}), which for $c(r)=1$ yields 
\begin{eqnarray}
\left[A\dot{A}^2\right]^{\prime}+2A^{\prime}\dot{A}^2=0,
\end{eqnarray}
which can be rewritten as 
\begin{eqnarray}
3\left[A\dot{A}^2\right]^{\prime}-2A\left[\dot{A}^2\right]^{\prime}=0\label{eqa1}\;.
\end{eqnarray}

A solution of  (\ref{eqa1}) is given by 
\begin{equation}
A(r,t)=\left[k_1(t)+k_2(r)\right]^{2/5}\label{a2}\;.
\end{equation}
Making the coordinate transformation  $x(r,t)=A(r,t)$, the line element  (\ref{ltb}) becomes 
\begin{equation}
dS^2=\left(1-\frac{4\dot{k}_1^2}{25x^3}\right)dt^2+\frac{4\dot{k}_1}{5x^{3/2}}dtdx-dx^2-x^2d\Omega^2\label{el1}\;,
\end{equation}
where $d\Omega^2=d\theta^2+\sin^2\theta d\phi^2$. Carrying out a new change of coordinates 
\begin{equation}
d\tau(x,t)=b_1(x,t)dt+b_2(x,t)dx\label{tc}\;,
\end{equation}
the line element (\ref{el1}) turns into
\begin{eqnarray}
dS^2&=&\frac{1}{b_1^2}\left(1-\frac{4\dot{k}_1^2}{25x^3}\right)d\tau^2-\left[\frac{2b_2}{b_1^2}\left(1-\frac{4\dot{k}_1^2}{25x^3}\right)-\frac{4\dot{k}_1}{5x^{3/2}b_1}\right]d\tau dx+\nonumber\\
&&-\left[1+\frac{4\dot{k}_1 b_2}{5x^{3/2}b_1}-\left(\frac{b_2}{b_1}\right)^2\left(1-\frac{4\dot{k}_1^2}{25x^3}\right)\right]dx^2-x^2d\Omega^2\label{el2}\;.
\end{eqnarray}
By using 
\begin{eqnarray}
b_1(r)=\sqrt{\frac{x^3-8M^3}{x^2(x-2M)}}\;,\;b_2(r)=2\sqrt{\frac{2M^3x}{(x-2M)(x^3-8M^3)}}\;,
\end{eqnarray}
and imposing 
\begin{eqnarray}
g_{\tau\tau}\equiv 1-\frac{2M}{x}\;,\;g_{\tau x}\equiv 0\;,\;k_1(t)\equiv 5\sqrt{2M^3}t\;,\label{imp1}
\end{eqnarray}
the line element  (\ref{el2}) reads 
\begin{eqnarray}
dS^2=\left(1-\frac{2M}{x}\right)d\tau^2-\left[1-\left(\frac{2M}{x}\right)^3\right]^{-1}dx^2-x^2d\Omega^2\label{el3}\;.
\end{eqnarray} 

The expression  (\ref{el3}) is a black hole solution of type-Schwarzschild, with the mass $M$ and horizon in  $x_H=2M$, but the Hawking temperature defined as $T_H=\sqrt{3}/8\pi M>T_{Schwarzschild}$. The usual case of Lemaitre-Tolman-Bondi, the solution is exactly that of Schwarzschild. Then, if we consider the increase of the mass as seen in the previous subsection, $M=5M_D/3$, we get the inequality  $T_H=\left(3\sqrt{3}/5\right)T_{HD}>T_{HD}=(1/8\pi M_{D})$, where $T_{HD}$ and  $M_D$ are the temperature and the mass in the diagonal case.


\subsection{Black hole in a dust-dominated universe}\label{sec5.3}

We can consider the results of the two previous subsections and generalize a solution for the black hole immersed in a dust-dominated universe. We proceed as follows. We need a solution such that when the mass $M$ is identically null, the solution corresponding to a universe dominated by the dust is recovered, while when we put  $\bar{\rho}_0=0$, the solution characterizes a black hole of type-Schwarzschild. Thus, just consider the simple linear combination
\begin{eqnarray}  
A(r,t)=rM\left(1+5\sqrt{2M^3}tr^{-5/2}\right)^{2/5}+\bar{\rho}_0
r\left(1+\sqrt{6\pi\bar{\rho}_0}t\right)^{2/3}\;,
\end{eqnarray}
which represents a black hole in a dust-dominated universe. The energy density can be easily calculated by the equation (\ref{dens2}), but we do not present this step due to its too long form. However, the limit $r,t\rightarrow 0$, leads to an infinite energy density, as at the Big Bang, and $t\rightarrow t_0$ leads to the current energy density, $\tilde{\rho}_0$. From the equation (\ref{dens2}), isolating $\rho$, we see that the energy density is always positive, for $t\geq -\sqrt{6\pi \bar{\rho}_0}$ and $r\geq 0$.


\subsection{Other exact solutions}\label{sec5.4}

Let us return to the system  (\ref{emnon}). This system contents 5 equations for 6 unknown functions. To solve it, we need one more additional equation. In any case, the system  (\ref{emnon}) has the very complicated form. So that the  finding its solutions is very hard job. For that reason let us simplify the task considering some particular cases. 
\subsubsection{Static  solution}

Here we now assume that $A=A(r)$ and $B=B(r)$ that corresponds to the static case.  Then the system  (\ref{emnon}) takes the form
\begin{eqnarray}\label{emnona}
\left\{\begin{array}{llll}
\frac{(B-A^{\prime})}{AB^2}T^{\prime}f_{TT}+f_{T}\left[\frac{T}{2}+\frac{1}{A^2}+\frac{A^{\prime}B^{\prime}}{AB^3}-\frac{A^{\prime}}{A^2B}-\frac{A^{\prime\prime}}{AB^2}\right]+\frac{f}{4}=4\pi\rho\,,\\
f_{T}\left[\frac{T}{2}+\frac{1}{A^2}-\frac{A^{\prime}}{A^2B}\right]+\frac{f}{4}=-4\pi p_{r}\,,\\
\frac{B-A^{\prime}}{2AB^2}T^{\prime}f_{TT}+\frac{f_{T}}{2}\left[\frac{A^{\prime}B^{\prime}}{AB^3}-\frac{A^{\prime\prime}}{AB^2}\right]+\frac{f}{4}=-4\pi p_{t}\;.
\end{array}\right.
\end{eqnarray}
 Note that in this case
\begin{eqnarray}\label{tenonb}
T=2\left[\frac{1}{A^2}-\frac{2A^{\prime}}{A^2B}+\left(\frac{A^{\prime}}{AB}\right)^2\right]\;.
\end{eqnarray}
From  (\ref{emnona}) we get
\begin{eqnarray}\label{emnonc}
\left\{\begin{array}{llll}
f_{T}\left[\frac{T}{2}+\frac{1}{A^2}-\frac{A^{\prime}}{A^2B}\right]-\frac{f}{4}=4\pi(\rho+2p_{t})\,,\\
f_{T}\left[\frac{T}{2}+\frac{1}{A^2}-\frac{A^{\prime}}{A^2B}\right]+\frac{f}{4}=-4\pi p_{r}\,,\\
\frac{B-A^{\prime}}{2AB^2}T^{\prime}f_{TT}+\frac{f_{T}}{2}\left[\frac{A^{\prime}B^{\prime}}{AB^3}-\frac{A^{\prime\prime}}{AB^2}\right]+\frac{f}{4}=-4\pi p_{t}\;.
\end{array}\right.
\end{eqnarray}
From the first two equations of this system  (\ref{emnonc}) we obtain
\begin{equation}\label{emnone}
f=-8\pi (\rho+2p_{t}+p_{r}).
\end{equation}

Finally we note that the system (\ref{emnonc}) contents  3 equations for 6 unknown functions ($A, B, f, \rho, p_t, p_r$). So to solve this system we need 3 additional equations. In the particular case of the vacuum, for $f(T)=\sum^{N}_{n=1}a_{n}T^n$, with $a_0=0$ ( without cosmological constant), one gets $f(T)=0$, implying that $T=0$. By integrating (\ref{tenonb}) for null torsion, we obtain $A^{\prime}(r)=B(r)$. Making a change of coordinates $R=A(r)$, the Minkowski's space metric is recovered in (\ref{ltb}). This result is consistent with the consideration of the equations in vacuum, as shown in \cite{ferraro}.  
\subsubsection{Time dependent   solution}

Now we try to get  a time dependent solution of the system (\ref{emnon}). From the second and third equations of (\ref{emnon}) we get
\begin{equation}\label{k1}
f_{T}=D_1\exp\left[\int dT\frac{(A^{\prime}\dot{B}-\dot{A}^{\prime}B)B}{(B-A^{\prime})\dot{T}}\right]
\end{equation}
and
\begin{equation}\label{k2}
f_{T}=D_2\exp{\left[-\int dT\frac{A^{\prime}\dot{B}-\dot{A}^{\prime}B}{B^3\dot{A}T^{\prime}}\right]},
\end{equation}
respectively. Here $D_i$ $ (i=1,2)$ are integration constants. Hence we come to the following constraint for the metric that is for the functions $A,B$:
\begin{equation}\label{k3}
\int dT\frac{(A^{\prime}\dot{B}-\dot{A}^{\prime}B)B}{(B-A^{\prime})\dot{T}}=D_3-\int dT\frac{A^{\prime}\dot{B}-\dot{A}^{\prime}B}{B^3\dot{A}T^{\prime}},
\end{equation}
where $D_3=\ln\frac{D_2}{D_1}$. Note that if $D_1=D_2$, this constraint takes the form
\begin{equation}\label{k3a}
B^4\dot{A}T^{\prime}=-(B-A^{\prime})\dot{T}.
\end{equation}
Note that Eqs. (\ref{k1})-(\ref{k2}) tell us that the function $f$ in terms of the metric $A,B$  expressed as
\begin{equation}\label{k4}
f=D_3+D_1\int dT\exp\left[\int dT\frac{(A^{\prime}\dot{B}-\dot{A}^{\prime}B)B}{(B-A^{\prime})\dot{T}}\right]
\end{equation}
or
\begin{equation}\label{k5}
f=D_4+D_2\int dT\exp{\left[-\int dT\frac{A^{\prime}\dot{B}-\dot{A}^{\prime}B}{B^3\dot{A}T^{\prime}}\right]},
\end{equation}
respectively.  Here $D_3$ and $D_4$ are integration constants. Now our aim is to express $\rho, p_r, p_t$ in terms of the metric functions. To do it, we rewrite the system (\ref{emnon}) as
\begin{eqnarray}\label{k6}
\left\{\begin{array}{llll}
K_2f_{TT}+K_1f_{T}+\frac{f}{4}=4\pi\rho,\\
-\frac{(B-A^{\prime})}{AB^2}\dot{T}f_{TT}+\frac{f_{T}}{AB}\left(A^{\prime}\dot{B}-\dot{A}^{\prime}B\right)=0,\\
-\frac{\dot{A}}{A}T^{\prime}f_{TT}-\frac{f_{T}}{AB^3}\left(A^{\prime}\dot{B}-\dot{A}^{\prime}B\right)=0,\\
N_2f_{TT}+N_1f_{T}+\frac{f}{4}=-4\pi p_{r},\\
M_2f_{TT}+M_1f_{T}+\frac{f}{4}=-4\pi p_{t}.
\end{array}\right.
\end{eqnarray}
where
\begin{eqnarray}\label{k7}
\left\{\begin{array}{lll}
K_2=\frac{(B-A^{\prime})}{AB^2}T^{\prime}, \quad K_1=\left[\frac{T}{2}+\frac{1}{A^2}+4\frac{\dot{A}\dot{B}}{AB}+\frac{A^{\prime}B^{\prime}}{AB^3}-\frac{A^{\prime}}{A^2B}-\frac{A^{\prime\prime}}{AB^2}\right],\\
N_2=\frac{\dot{A}}{A}T^{\prime},\quad N_1=\left[\frac{T}{2}+\frac{\ddot{A}}{A}+3\frac{\dot{A}\dot{B}}{AB}+\frac{1}{A^2}-\frac{A^{\prime}}{A^2B}\right],\\
M_2=\frac{1}{2}\left[\left(\frac{\dot{A}}{A}+\frac{\dot{B}}{B}\right)\dot{T}+\frac{B-A^{\prime}}{AB^2}T^{\prime}\right], \quad M_1=\frac{1}{2}\left[\frac{T}{2}+\frac{\ddot{A}}{A}+\frac{\ddot{B}}{B}+5\frac{\dot{A}\dot{B}}{AB}+\frac{A^{\prime}B^{\prime}}{AB^3}-\frac{A^{\prime\prime}}{AB^2}\right].
\end{array}\right.
\end{eqnarray}
Let us eliminate $f_{TT}$ from (\ref{k6}). To do it, we can use  the second or third equations of the system (\ref{k6}). As result we come to the equations
\begin{equation}\label{k8}
f_{TT}=\frac{(A^{\prime}\dot{B}-\dot{A}^{\prime}B)B}{(B-A^{\prime})\dot{T}}f_{T}\equiv L_1f_{T}
\end{equation}
and
\begin{equation}\label{k9}
f_{TT}=-\frac{A^{\prime}\dot{B}-\dot{A}^{\prime}B}{B^3\dot{A}T^{\prime}}f_{T}\equiv L_2f_{T}.
\end{equation}
Note that from these two equations follows that $L_1=L_2$ that is equivalent to the constraint (\ref{k3a}).  Using the equations (\ref{k8}) or (\ref{k9}), from (\ref{k6}) we get
\begin{eqnarray}\label{k10}
\left\{\begin{array}{llll}
(K_2L_1+K_1)f_{T}+\frac{f}{4}=4\pi\rho\,,\\
(N_2L_1+N_1)f_{T}+\frac{f}{4}=-4\pi p_{r}\,,\\
(M_2L_1+M_1)f_{T}+\frac{f}{4}=-4\pi p_{t}\;.
\end{array}\right.
\end{eqnarray}
Hence finding for example $f_T$ as
\begin{eqnarray}\label{k11}
f_T=\frac{4\pi \rho -f/4}{K_2L_1+K_1},
\end{eqnarray}
we finally come to the following formula for $f(T)$:
\begin{eqnarray}\label{k12}
f=-16\pi \left(1-\frac{N_2L_1+N_1}{K_2L_1+K_1}\right)^{-1}\left(\frac{N_2L_1+N_1}{K_2L_1+K_1}\rho+p_r\right)
\end{eqnarray}
or
\begin{eqnarray}\label{k13}
f=16\pi\left(3-W\right)^{-1}\left[\left(1-W\right)\rho-p_r-p_t\right],
\end{eqnarray}
where
\begin{eqnarray}\label{k14}
W=\frac{(K_2+N_2+M_2)L_1+K_1+N_1+M_1}{K_2L_1+K_1}.
\end{eqnarray}

Once again we have here an algebraic function $f(T)$ which depends on the matter content in (\ref{k13}). This confirms again that a possibility of getting a consistency of this theory is considering a matter content depending on the algebraic function $f(T)$ and its derivatives, as shown in \cite{stephane2}.
\section{Conclusion}\label{sec6}
\hspace{0,2cm} We obtained the equations of motion for the $f(T)$ theory  in (\ref{em}). We first took a set of diagonal tetrads for the case of the Lema\^itre-Tolman-Bondi's (LTB) metric and obtained the same results as that of the General Relativity (GR) in (\ref{em2}). This does not seem surprising since it is well known that the Teleparallel Theory is dynamically equivalent to the GR \cite{pereira}, which is particularly our case here. We explored a particular case of an universe dominated by the matter for comparing it with the case of a set of non-diagonal tetrads.
\par
Afterwards, we chose a new set of non-diagonal tetrads for projecting the metric of LTB in the tangent space and obtained new equations of motion of this case. This result, that the $f(T)$ theory  possesses a dependence on the frame in its description \cite{yapiskan}, also is not surprising, and the fact that the equations in the non-diagonal frame being different from that of the diagonal one was already expected. We explained the same example of an universe dominated by the matter and we noted that the increased (decreased) in the mass (or energy density) is possibly dependent on the time, what is drastically different from the GR. We also perform the example of a black hole solution, which is of type-Schwarzschild and a slightly higher Hawking temperature. Our last example is that of black hole in a dust-dominated universe, which produces the same result as in the case of GR.
\par
Through a set of non-diagonal tetrads we still were able to make various analysis already made in the GR, as the evolution of the black holes apparent horizon (AH) and cosmic AH \cite{gao}, CMB \cite{cmb,alnes} and many other possibilities, but which will be minutely addressed in a future work. Hence, we make possible the analysis of the other usual cosmological and astrophysical phenomena, already realized in the GR, but which still have some obscure  points to be explained.


\vspace{0,25cm}
{\bf Acknowledgement:}  Authors thank E. N. Saridakis for useful discussions.  M. H. Daouda thanks CNPq/TWAS for financial support. M. E. Rodrigues is grateful to  UFES and UFPA for the hospitality during the development of this work and thanks CNPq for financial support. M. J. S. Houndjo thanks CNPq/FAPES for financial support.

\end{document}